\documentclass[a4paper,11pt]{article}
\usepackage{pos}

\title{The VERITAS-Stellar Intensity Interferometry (VSII) survey of Stellar Diameters}
 \ShortTitle{VERITAS Stellar Diameter Survey}

\author*[a]{David Kieda}
\author[a,b]{Jonathan Davis}
\author[a]{Tugdual LeBohec}
\author[c]{Mike Lisa}
\author[a,d]{Nolan K. Matthews}
\affiliation[a]{University of Utah,\\
  Department of Physics and Astronomy, 115 S 1400 E \#201, Salt Lake City, Utah, USA }
   \affiliation[b]{current address: Cornell University,\\
  Department of Astronomy, Ithaca, NY,  USA }
    \affiliation[c]{The Ohio State University,\\
  Department of Physics, Columbus, OH, USA }
    \affiliation[d]{current address: Universit\'{e} Côte d'Azur, CNRS, INPHYNI, France}  
  
\forColl{VERITAS} 

\emailAdd{dave.kieda@utah.edu}

\abstract{The VERITAS Imaging Air Cherenkov Telescope (IACT) array was augmented in 2019 with high-speed focal plane electronics to allow its use for Stellar Intensity Interferometry (SII) observations. Since January 2019, the VERITAS Stellar Interferometer (VSII) recorded more than 250 hours of moonlit observations on 39 different bright stars and binary systems ($m_V < 3.74$) at an effective optical wavelength of 416 nm. These observations resulted in the measurement of the diameters of several stars with better than 5\% resolution. This talk will describe the status of the VSII survey and  analysis.}

\FullConference{37$^{\rm{th}}$ International Cosmic Ray Conference (ICRC 2021)\\
		July 12th -- 23rd, 2021\\
		Online -- Berlin, Germany}

\begin{document}
\maketitle

\section{Introduction}

The SII technique measures correlated fluctuations in starlight intensity between spatially separated telescopes.  The VERITAS Stellar Intensity Interferometer (VSII)  is currently conducting a comprehensive survey of  northern sky bright stars, with the goal of  measuring the diameters of 30+ stellar objects to an precision of 5\% or better.  the VSII survey also allows a quantitative exploration of the sensitivity of the VSII observatory and has led to continuous improvements in both instrumentation and analysis techniques.  An accompanying paper at this conference  \cite{Kieda2021} describes the current status of the VSII observatory, including recent upgrades and improvements motivated by the initial  results  of the VSII survey observations.  

\section{VSII Nightly Observation Procedure}
Nightly SII observations begin about 1 hour before twilight. The observer opens the shutter to each VSII camera, and deploys a collapsable mirror  to a 45-degree position to reflect the starlight onto a photomultiplier tube (PMT) mounted transverse to the optical axis. The PMT High Voltage (HV) supply battery is reconnected after daytime charging, and the HV supply is powered up.  The Data Acquisition (DAQ)  system for each telescope  is then started  and short (10-30 second duration) calibration runs (with HV = 0V) are performed  with the telescopes parked in their stow positions. The pedestals of the data traces are recorded and used for removing the DC offset in each electronics channel. About 30 minutes before the start of the observation, the HV is turned on to a sufficient level to verify that each PMT is operational and that the telescopes are ready for beginning observations. The telescopes are then slewed to the first target of the night, and the CCD camera for each telescope is used to micro-adjust the pointing of the telescope to center the point spread function of the starlight onto the narrowband filter.   Finally, the HV in each telescope is adjusted to provide a nominal DC current  in each PMT (about 10 microamps for a $m_V = 2$ star).

Once the PMT currents are equalized, the telescopes are slewed $0.5^\circ$ off from the star's sky position, and a 1 minute OFF run is taken to measure the night sky background near the star's sky field. The telescope is then slewed to point at the target star, and a 30-90 minute continuous observation is performed (an ON run).  Each observation  run is triggered at every telescope by a central trigger pulse that is distributed to each DAQ system via a fiber optic system.   As the VERITAS telescopes track the target over 4-6 hours, the projected distance between the telescope combinations continuously changes with the movement of the star position in elevation and azimuth.  The measurement of the correlated fluctuations between two telescopes (interferometric visibility  $|V(r)|^2$) is therefore is sampled over a range of projected distances $r$ during a single night’s
observations.

During the ON run, the point spread function (PSF) of the starlight on the focal plane is
tracked. If the PSF moves off the narrowband filter window, manual telescope tracking adjustments
are performed to recenter the starlight on the light sensor. At the end of the ON run, another
OFF run is taken with a $0.5^\circ$  offset, and the ON-OFF observing sequence is repeated, as necessary.
Moving to a new observing target requires slewing to the desired target, making the manual tracking
adjustments to center the PSF onto the optical filter, setting the HV to equalize the PMT gains, and
then performing the nominal OFF-ON-OFF run sequence.

At the end of the observing night the HV is turned off and the telescopes are brought to their stow positions.  The HV battery is connected to the battery charger, and the collapsible mirror is folded down. The camera shutter is then closed to protect the focal plane during the daytime.

\section{VSII Observing Target Selection }
The suitability of a target for any given night of observation depends upon several factors. These factors include source-specific 
characteristics (stellar classification and magnitude, stellar diameter, single/binary/multiple star configuration),  observatory
characteristics (observatory latitude/longitude, telescope diameters, telescope separations, electronic noise and optical efficiencies), and
astronomical considerations  (Source RA/Dec, number of hours observable on a given night, seasonal nighttime visibility of the source). 
VSII uses the public domain ASIIP software package \cite{davis,ASIIP}  to list and rank the suitability of potential sources for any given observation night.

\begin{figure}[!h]
  \vspace{4mm}
  \centering
  \includegraphics[width=4.0in]{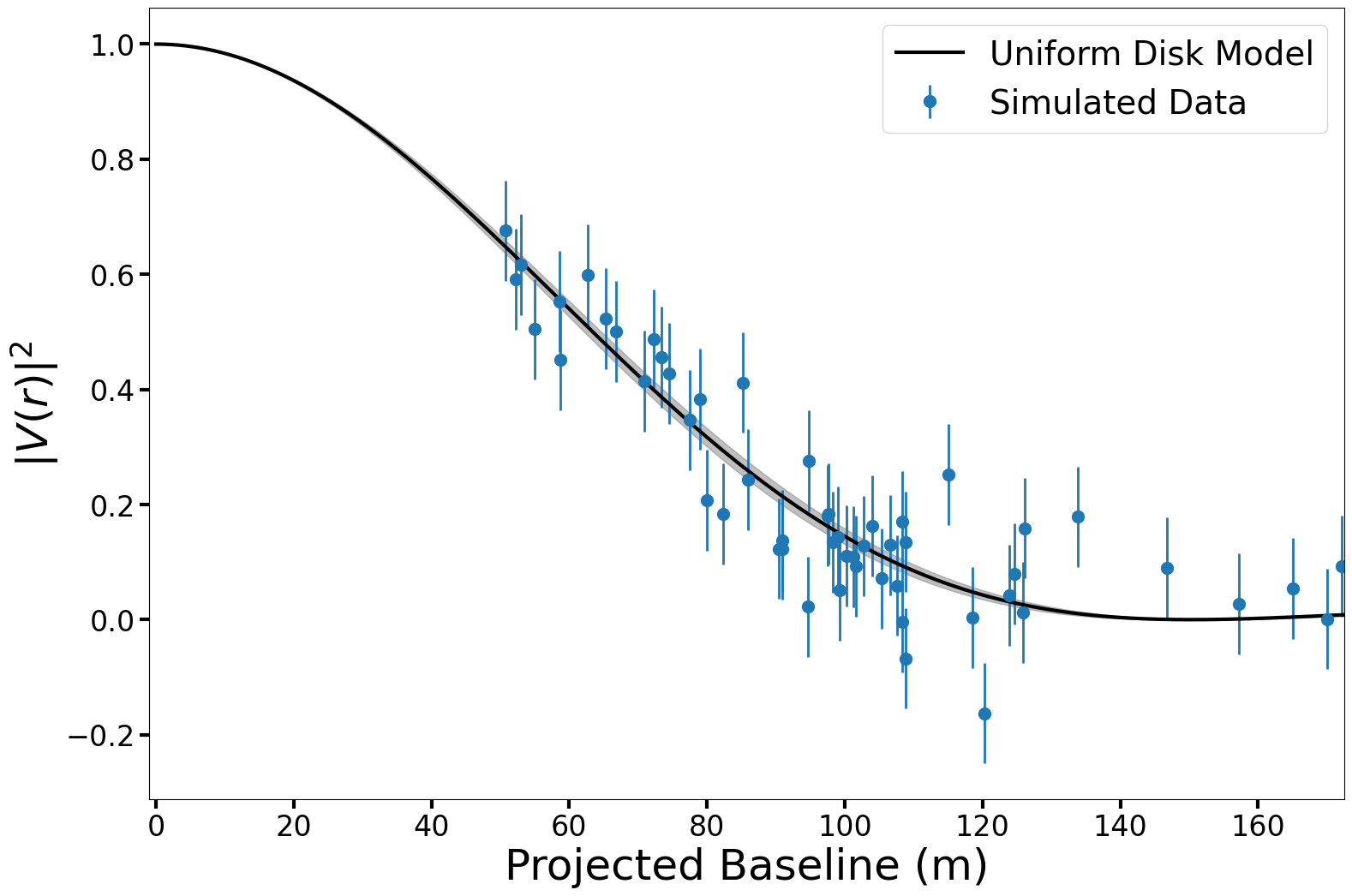}
  \caption{ASIIP simulation of VSII observation of Bellatrix ( $\gamma$ {\it Ori}) for the night of February 27, 2021. The individual data points represent simulated visibility measurements for Bellatrix,  given its trajectory across the night sky, VERITAS telescope locations, and astronomical twilight constraints. The ASIIP simulation estimates a 2.0\% error in the determination of the angular diameter. The simulation does not include the effects of the moonlight background on the observation (The moon angle on this night ranged from $41.5^\circ$ to $42^\circ$). Vertical Axis: Normalized two-telescope cross-correlation (interferometric visibility$|V(r)|^2$  measured
by a telescope pair at a baseline $r$. Horizontal axis: projected baseline separation between telescope pair ($m$).  }
 \label{Figure1}
 \end{figure}

ASIIP simulations are performed for a specified SII observing week during the year, and potential observation targets (drawn from several catalogs, including the JMMC catalog \cite{JMMC}) are ranked according to the estimated error in the determination of the stellar diameter.  For examples, large  diameter  stars ($>1\ mas$) may be poorly fit of the source can only be observed directly overhead on a given night. In contrast, smaller stars ($< 0.5\ mas$) may be unresolved if the observations are constrained to  low elevations on a given night. 
At the beginning of the observing night, the simulated visibility curves for each target (e.g. Figure \ref{Figure1} ) are reviewed, and an observing sequence of 2-5 stellar observations is generated.  Then, the observing plan is sequenced to provide an appropriate set of observation hours and baselines for each target to result in a  quality measurement of each visibility curve.  

 \subsection{Effects of Moon Angle on Observational Planning}
 
 The narrowband Semrock  interference filter  (5 $nm$ bandpass, normal incidence)  \cite{semrock} substantially reduces the intensity of background light impinging on the VSII photomultipler  tubes (PMTs).   This allows VSII observation to occur at all moon phases, including the full moon. In practice, the moonlight places additional restrictions on the observability of  specific stars on a given night. The effects include:

\begin{enumerate}
\item Moonlight shining directly on the focal plane. Direct moonlight on the focal plane makes it difficult to view the location of the starlight PSF  on the interference filter aperture using the focal plane CCD camera.  The direct moonlight makes it challenging to visualize the tracking adjustments necessary to keep the starlight PSF centered on the narrowband filter.  To first order, this issue rules out any observations when the moon angle to the target source is  $> 90^\circ$. In practice, the VSII focal plane is slightly recessed into the VERITAS camera body, providing additional baffling against the moonlight. This baffling allows targets to be observed with moon angles $<95^\circ$. 

\item Moonlight is scattered by the atmosphere through both Rayleigh and Mie Scattering. The scattering functions  are peaked in the forward direction, creating a halo of scattered moonlight around the moon. SII observations are dchallenging when the target star is less than $30^\circ$  aways from the moon position. Moon  angles less than $30^\circ$  also make it challenging to make tracking adjustments. 

\item Atmospheric conditions can cause additional constraints on the observability of targets. Patchy clouds in the night sky can strongly scatter moonlight onto the focal plane at unanticipated angles, making it difficult to  perform reliable tracking or estimate night sky background biases in the visibility curves.  Low clouds on the horizon can restrict observation to higher elevations (longer baselines), making it difficult to sample the central peak of the visibility curve for  stars with diameters greater  than  1 mas. Alternatively, atmospheric conditions that restrict a given night's observations to  low elevations (short baselines) will result in smaller stars being unresolved.  

\item Observing stars at low elevation will noticeably dim their magnitude due to the increased atmospheric depth.  If moon angle constraints force  observations to occur only at low elevation angles, visible dimmer magnitude targets ( $m_V>2.5$ ) may not be feasible observations.

\item Dimmer stars ($m_V>3.0$ ) are difficult to track under most moonlight conditions, regardless of the moon angle. therefore, the nightly observation plan must schedule time between astronomical twilight and moonrise/moonset to perform such observations. 

 \item In addition, a larger amount background light compared to the starlight will bias the visibility curves, and result in a systematic error in determining the stellar radius.
The night sky background light correction is most straightforward when the background light is less intense, and the background does not change significantly across short angular distances across the night sky.
\end{enumerate}

 \begin{figure}[!h]
  \vspace{5mm}
  \centering
  \includegraphics[width=4.in]{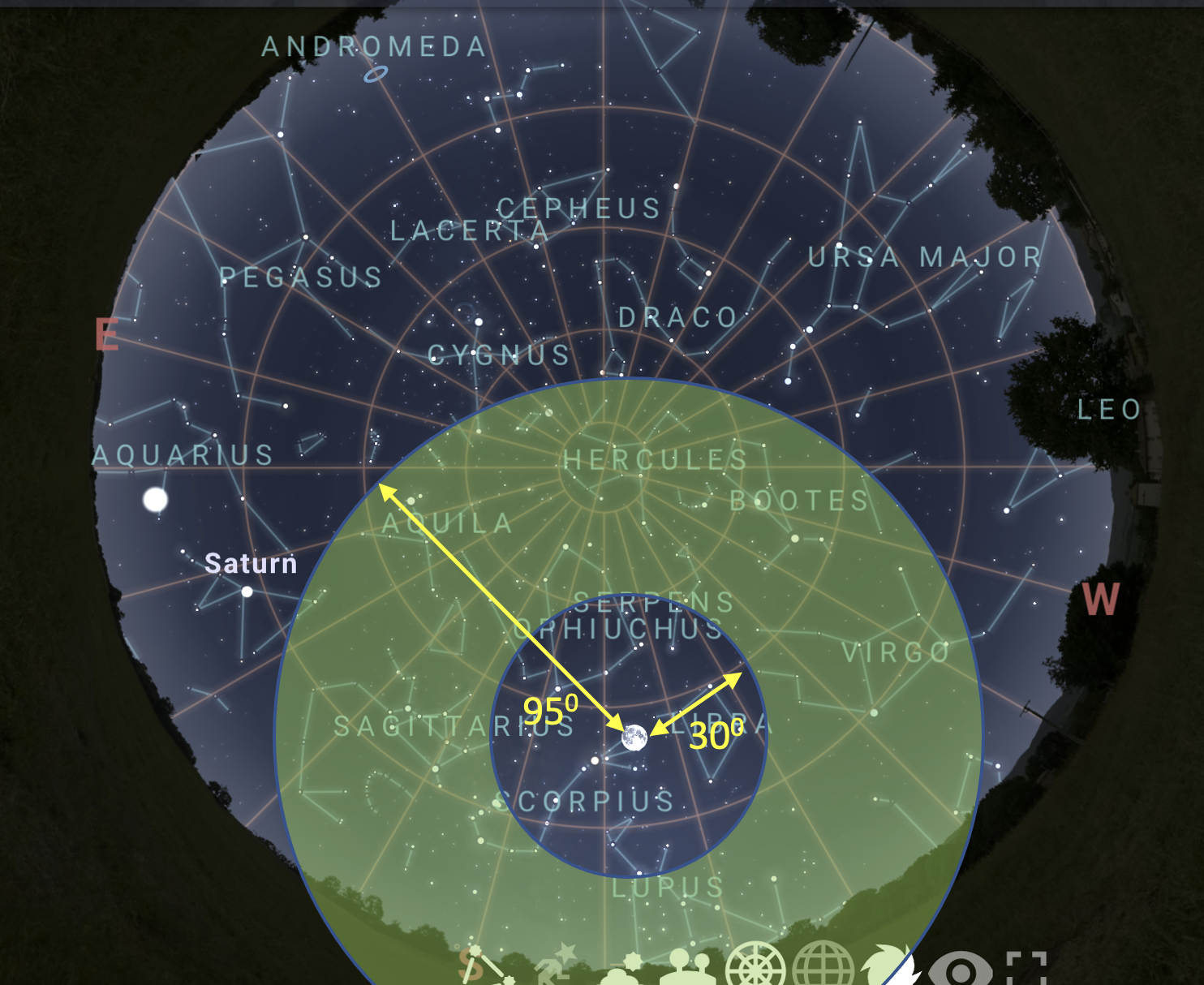}
  \caption{Moonlight constrained target selection for a typical SII observation night.  Feasible source observations will have moon angles between $30^\circ$ and $95^\circ$ .
The range of viable source observations for the selected time and date lie within the yellow annulus  region of the sky. This constraint severely limits the number  of observable sources on any given night. Because the moon moves $\approx 13^\circ$ through the  night sky each day,  the list of potential source observations changes daily.  The All-sky constellation plot was generated using Stellarium \cite{stellarium}.   }
  \label{Figure2}
 \end{figure}
 
 At the beginning of the observing week, the first target observed is chosen to
  be a bright SII reference (previously observed) star ($m_V < 2.2$) with
 $0.6\ mas < stellar \ radius < 0.85\  mas $   to verify the basic operational status of VSII.
 The star must have an appropriate moon angle ($30^\circ < moon \ angle < 95^\circ$) and the star is selected to give a strong 
 visibility signal with a 1 to 1.5  hour exposure.  
 
 Priority for  Subsequent observations is evaluated using the following priorities:
\begin{enumerate} 
\item $30^\circ < moon \ angle < 95^\circ$
\item Observation time  $> 1 $ hour
\item $0.4 \ mas < stellar \ diameter < 1.2 \ mas$
\item Quality of ASIIP constraint on stellar diameter
\item Prefer $ m_V < 3 $
\item Prefer O, then B, then A stars
\item Previously unobserved  targets have a priority
\item Underexposed targets have a priority
\item Short period orbital binaries (e.g. Spica)  establish  a multi-day priority  to map out a visibility curve at different phases of the orbit. 
\item Unusual stellar characteristics (e.g. cepheid, fast rotators, etc) gain priority over `vanilla'-type singular stars.
\end{enumerate}

\begin{figure}[!h]
  \vspace{5mm}
  \centering
  \includegraphics[width=6.in]{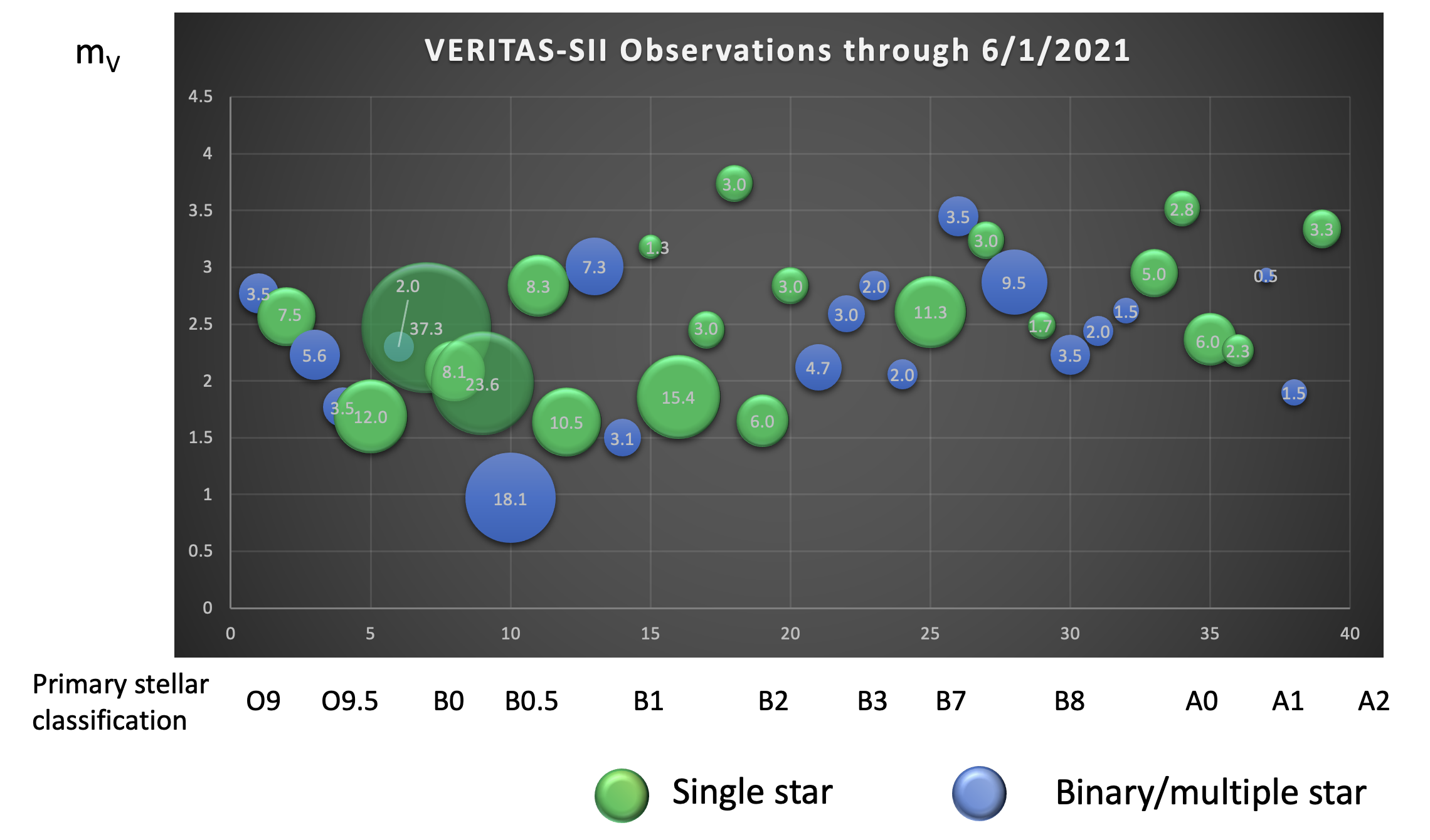}
  \caption{Summary of VSII survey  observations from December 2019 through June 1, 2021. Horizontal Axis: Primary Stellar classification. Vertical Axis: Visual magnitude $m_V$.  Green Circles: Single Star observations. Blue Circles: Binary/multiple star  system observations. For each source, the area of each circle is proportional to the number of hours observed. The numbers inside each circle indicates the  number of  hours observed for that source. }
  \label{Figure3}
 \end{figure}
 
    \section{Summary of VSII Survey Observations} 
 
 Since December 2019, VSII has performed more that 250 hours of SII observation on 39 different astronomical targets (Figure \ref{Figure3}). The survey includes  21 single stars and 18 binary/multiple star systems ranging in magnitude from $ 0.97 < m_V < 3.74$.  The primary stellar classification ranges from O9 through A2 . Figure \ref{Figure3} illustrates the number of observation hours on each target  as a function of stellar classification (temperature), stellar magnitude $m_V$, and single/binary star system classification.  This plot include VSII observations through June 1, 2021. The observation exposure is weighted towards longer observation hours for bright ($m_V< 2.5$) O/B0/B1 stars, but there are a substantial number of observation hours spent probing the sensitivity of VSII to dimmer B7/B8/A stars ($2.5 < m_V < 3.5$).  Both single stars and multiple stars are broadly represented across visual magnitude and stellar classification.

    \section{Data Analysis and Extraction of Visibility Curves} 
\begin{figure}[!h]
  \vspace{5mm}
  \centering
  \includegraphics[width=5.in]{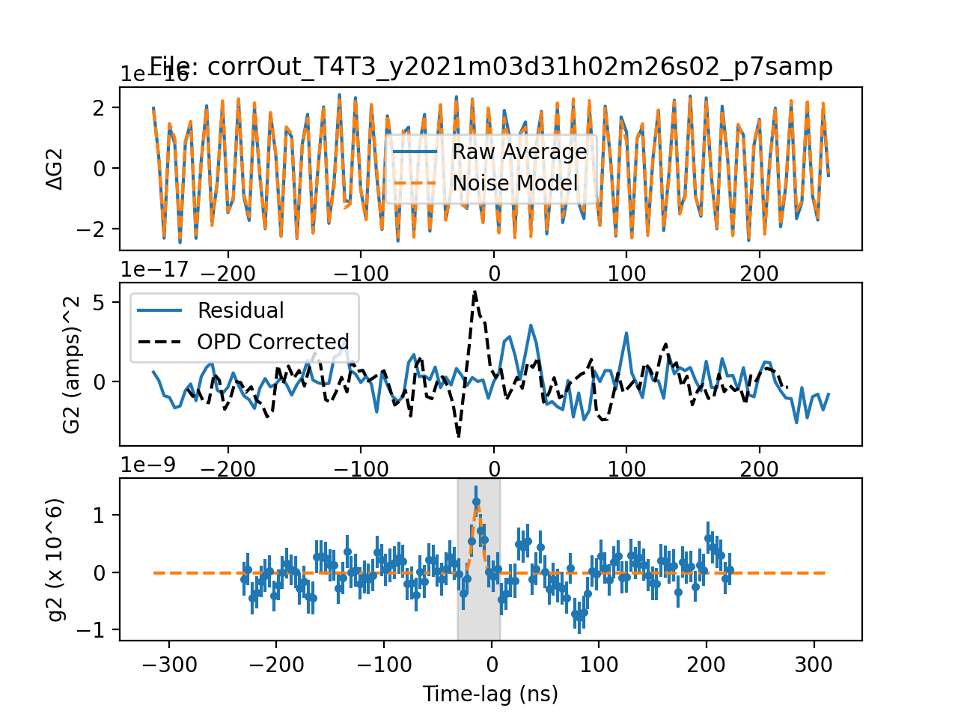}
  \caption{Analysis of a raw two-telescope correllelogram for the extraction of the visibility at a specific telescope separation. The specific observation took place on March 31, 2021. The correllelogram measures the correlation between fluctuations in light intensities  recorded by VSII telescope 3 and 4 as a function of the time-lag between the photon arrival at each telescope. Upper panel: Fitting the raw correllelogram to remove RF noise pickup. Middle panel: Residual correlation signal before and after correction for optical path delay. Bottom panel: Fitting the OPD corrected correllogram to a Gaussian model to extract the magnitude of the two-telescope correlation (visibility). Horizontal axis : time-lag between telescope data streams (ns). Vertical axis: scaled interferometric visibility (scaling of units depend upon each panel)  }
  \label{Figure4}
 \end{figure}

   Raw data from individual source observations are processed on-site into a "correllelogram"  using a pipelined two-telescope cross-correlation algorithm that  the FGPAs host  in the VSII Data Acquisition crates. Further details regarding the cross-correlation analysis is provided in a separate paper at this conference \cite{Kieda2021}.  Once the correllelogram for each two-telescope combination is computed, each correllelogram  is analyzed to extract the magnitude of the visibility at the specific telescope separation. Figure \ref{Figure4} illustrates the analysis steps of each two-telescope  correllelogram. This procedure is complicated by the presence of RF 79 Mhz noise in the raw  data stream.
   
   First,  the raw correllogram is fitted for a noise model, including a dominant 79 MHz component and several side frequencies. The weighting of each component is iteratively adjusted to match the raw correllogram data until the values converge on a satisfactory fit  (Figure \ref{Figure4}, upper panel). Next, the residual between the noise model and the raw data is calculated, and the residual data is corrected for the changing optical path delay during the  observation using the known projected distance difference between the two telescope to the target   (Figure \ref{Figure4}, middle panel). Finally a Gaussian fit is used to extract the peak of the visibility curve at the expected time lag between telescopes (Figure \ref{Figure4}, bottom panel).  A visibility curve similar to Figure \ref{Figure1} is then calculated using the measured visibility peaks and known telescope separations for every telescope pair in the observation. The final visibility curve must be corrected for the presence of night sky background. After this correction, a suitable stellar diameter model is  fitted to the visibility curve to extract a reliable measurement of the stellar diameter \cite{SII2020,Matthews2020}.  We will present a sampling of preliminary stellar diameter measurements for selected stars from the VSII survey at the conference.  
 
 \section{Acknowledgements}
This research is supported by grants from the US Department of Energy Office of Science, the US National Science Foundation, the Smithsonian Institution, and NSERC in Canada. This research used resources provided by the Open Science Grid, which is supported by the National Science Foundation and the US Department of Energy's Office of Science, and resources of the National Energy Research Scientific Computing Center (NERSC), a US Department of Energy Office of Science User Facility operated under Contract No. DE-AC02-05CH11231. The authors gratefully acknowledge support under NSF Grant \#AST 1806262 for the fabrication and commissioning of the VERITAS-SII instrumentation. We acknowledge the excellent work of the technical support staff at the Fred Lawrence Whipple Observatory and the collaborating institutions in the construction and operation of the instrument.

 \clearpage
 \section*{Full Authors List: \Coll\ Collaboration}
 
 \scriptsize
 \noindent
 C.~B.~Adams$^{1}$,
 A.~Archer$^{2}$,
 W.~Benbow$^{3}$,
 A.~Brill$^{1}$,
 J.~H.~Buckley$^{4}$,
 M.~Capasso$^{5}$,
 J.~L.~Christiansen$^{6}$,
 A.~J.~Chromey$^{7}$, 
 M.~Errando$^{4}$,
 A.~Falcone$^{8}$,
 K.~A.~Farrell$^{9}$,
 Q.~Feng$^{5}$,
 G.~M.~Foote$^{10}$,
 L.~Fortson$^{11}$,
 A.~Furniss$^{12}$,
 A.~Gent$^{13}$,
 G.~H.~Gillanders$^{14}$,
 C.~Giuri$^{15}$,
 O.~Gueta$^{15}$,
 D.~Hanna$^{16}$,
 O.~Hervet$^{17}$,
 J.~Holder$^{10}$,
 B.~Hona$^{18}$,
 T.~B.~Humensky$^{1}$,
 W.~Jin$^{19}$,
 P.~Kaaret$^{20}$,
 M.~Kertzman$^{2}$,
 D.~Kieda$^{18}$,
 T.~K.~Kleiner$^{15}$,
 S.~Kumar$^{16}$,
 M.~J.~Lang$^{14}$,
 M.~Lundy$^{16}$,
 G.~Maier$^{15}$,
 C.~E~McGrath$^{9}$,
 P.~Moriarty$^{14}$,
 R.~Mukherjee$^{5}$,
 D.~Nieto$^{21}$,
 M.~Nievas-Rosillo$^{15}$,
 S.~O'Brien$^{16}$,
 R.~A.~Ong$^{22}$,
 A.~N.~Otte$^{13}$,
 S.~R. Patel$^{15}$,
 S.~Patel$^{20}$,
 K.~Pfrang$^{15}$,
 M.~Pohl$^{23,15}$,
 R.~R.~Prado$^{15}$,
 E.~Pueschel$^{15}$,
 J.~Quinn$^{9}$,
 K.~Ragan$^{16}$,
 P.~T.~Reynolds$^{24}$,
 D.~Ribeiro$^{1}$,
 E.~Roache$^{3}$,
 J.~L.~Ryan$^{22}$,
 I.~Sadeh$^{15}$,
 M.~Santander$^{19}$,
 G.~H.~Sembroski$^{25}$,
 R.~Shang$^{22}$,
 D.~Tak$^{15}$,
 V.~V.~Vassiliev$^{22}$,
 A.~Weinstein$^{7}$,
 D.~A.~Williams$^{17}$,
 and 
 T.~J.~Williamson$^{10}$\\
 \noindent
 $^{1}${Physics Department, Columbia University, New York, NY 10027, USA}
 $^{2}${Department of Physics and Astronomy, DePauw University, Greencastle, IN 46135-0037, USA}
 $^{3}${Center for Astrophysics $|$ Harvard \& Smithsonian, Cambridge, MA 02138, USA}
 $^{4}${Department of Physics, Washington University, St. Louis, MO 63130, USA}
 $^{5}${Department of Physics and Astronomy, Barnard College, Columbia University, NY 10027, USA}
 $^{6}${Physics Department, California Polytechnic State University, San Luis Obispo, CA 94307, USA} 
 $^{7}${Department of Physics and Astronomy, Iowa State University, Ames, IA 50011, USA}
 $^{8}${Department of Astronomy and Astrophysics, 525 Davey Lab, Pennsylvania State University, University Park, PA 16802, USA}
 $^{9}${School of Physics, University College Dublin, Belfield, Dublin 4, Ireland}
 $^{10}${Department of Physics and Astronomy and the Bartol Research Institute, University of Delaware, Newark, DE 19716, USA}
 $^{11}${School of Physics and Astronomy, University of Minnesota, Minneapolis, MN 55455, USA}
 $^{12}${Department of Physics, California State University - East Bay, Hayward, CA 94542, USA}
 $^{13}${School of Physics and Center for Relativistic Astrophysics, Georgia Institute of Technology, 837 State Street NW, Atlanta, GA 30332-0430}
 $^{14}${School of Physics, National University of Ireland Galway, University Road, Galway, Ireland}
 $^{15}${DESY, Platanenallee 6, 15738 Zeuthen, Germany}
 $^{16}${Physics Department, McGill University, Montreal, QC H3A 2T8, Canada}
 $^{17}${Santa Cruz Institute for Particle Physics and Department of Physics, University of California, Santa Cruz, CA 95064, USA}
 $^{18}${Department of Physics and Astronomy, University of Utah, Salt Lake City, UT 84112, USA}
 $^{19}${Department of Physics and Astronomy, University of Alabama, Tuscaloosa, AL 35487, USA}
 $^{20}${Department of Physics and Astronomy, University of Iowa, Van Allen Hall, Iowa City, IA 52242, USA}
 $^{21}${Institute of Particle and Cosmos Physics, Universidad Complutense de Madrid, 28040 Madrid, Spain}
 $^{22}${Department of Physics and Astronomy, University of California, Los Angeles, CA 90095, USA}
 $^{23}${Institute of Physics and Astronomy, University of Potsdam, 14476 Potsdam-Golm, Germany}
 $^{24}${Department of Physical Sciences, Munster Technological University, Bishopstown, Cork, T12 P928, Ireland}
 $^{25}${Department of Physics and Astronomy, Purdue University, West Lafayette, IN 47907, USA}

%
%
%


\begin{thebibliography}{99}
 
\bibitem{Kieda2021} Kieda, D. et al,  Status of the VERITAS Stellar Intensity Interferometry (VSII) System, {\em  Proc. 37th International Cosmic Ray Conference } POS(ICRC 2021) (this conference).   
\bibitem{davis} Davis, J., Matthews, N. and Kieda, D., ASIIP: A Stellar Intensity Interferometry Target Planner, J. Ast. Inst. 2020 {\bf 6}, 037001.
\bibitem{ASIIP} https://github.com/astronomaestro/ASIIP
\bibitem{JMMC} Duvert, G. et al.,  VizieR Online Data Catalog: JMDC : JMMC Measured Stellar Diameters Catalogue., VizieR Online Data Catalog 2345, November 2016.
\bibitem{semrock} https://www.semrock.com/filterdetails.aspx?id=ff01-420/5-25 
\bibitem{stellarium} https://stellarium-web.org/
\bibitem{SII2020} Abeysekara, A. U. et al. Demonstration of stellar intensity interferometry with the four VERITAS telescopes, Nature Astronomy 2020  https://www.nature.com/articles/s41550-020-1143-y 
\bibitem{Matthews2020} Matthews, N. , Intensity Interferometry Observations with VERITAS, PhD Dissertation, University of Utah (2020).
\end{thebibliography}
\end{document}